\documentclass[graybox]{svmult}

\usepackage{mathptmx}       
\usepackage{helvet}         
\usepackage{courier}        
\usepackage{type1cm}        
\usepackage{makeidx}         
\usepackage{graphicx}        
\usepackage{multicol}        
\usepackage[bottom]{footmisc}

\usepackage{amsmath}
\usepackage{amssymb}
\usepackage{amsfonts}
\usepackage{graphicx}
\usepackage{caption}
\usepackage[english]{babel}
\usepackage{floatrow}
\usepackage{float}
\usepackage{epsfig}
\usepackage{epstopdf}
\usepackage{grffile}
\usepackage{blkarray}
\usepackage {tikz}
\usepackage{multirow} 
\usepackage{rotating} 

\makeindex             

\begin{document}

\title*{High-order compact finite difference scheme for option pricing
  in stochastic volatility with contemporaneous jump models}
 \titlerunning{HOC finite difference scheme for option pricing
  in SVCJ models }
\author{Bertram D{\"u}ring and Alexander Pitkin}
\institute{Bertram D{\"u}ring and Alexander Pitkin \at Department of Mathematics, University of Sussex, Pevensey II, Brighton, BN1 9QH, United Kingdom, \email{bd80@sussex.ac.uk, a.h.pitkin@sussex.ac.uk}}
%
%
\maketitle

\abstract{ 
We extend the scheme developed
  in B.~D\"uring, A.~Pitkin, "High-order compact
 finite difference scheme for option pricing in stochastic volatility jump models", 2019, to the so-called stochastic volatility with
 contemporaneous jumps (SVCJ) model, derived by Duffie, Pan and Singleton.
The performance of the scheme is assessed through a number of numerical experiments, using comparisons against a standard second-order central difference scheme. We observe that the new high-order compact scheme achieves fourth order convergence and discuss the effects on efficiency and computation time.}


\section{Introduction}
\label{sec:1}

The stochastic volatility with contemporaneous jump model  (SVCJ) model, \cite{DuPaSi00}, can be seen as an extension of the Bates model \cite{Bates},
 which combines the positive features of stochastic volatility and jump-diffusion models.
In both models the option price is given as the solution of a partial integro-differential equation (PIDE),
see e.g.\ \cite{Cont}. In \cite{DuPi17} we have presented a new high-order compact finite
difference scheme for option pricing in Bates model. The
implicit-explicit scheme is based on the approaches in D\"uring and
Fourni\'e \cite{DuFo12} and Salmi {\em et al.\/}
\cite{Salmi14}. The scheme is fourth order accurate in space and second order accurate
in time. In the present work we extend the scheme to the SVCJ model derived by Duffie, Pan and Singleton \cite{DuPaSi00}. 


This article is organised as follows. In the next section we recall the SVCJ model for option pricing, we discuss the implementation of the implicit-explicit scheme and note the adaptations to the previously derived scheme for option pricing under the Bates model. Section~\ref{sec:3} is devoted to the numerical experiments, where we assess the performance of the new scheme. 

\section{The SVCJ Model}
\label{sec:2}

The SVCJ model \cite{DuPaSi00} is a stochastic volatility model which allows for jumps in both volatility and returns. Within this model the behaviour of the asset value, \textit{S}, and its variance, \( \sigma \), is described by the coupled stochastic differential equations,
\begin{align*}
dS(t) &= \mu_S S(t) dt + \sqrt{\sigma (t)} S(t) dW_1(t) + S(t) dJ^S,\\
d\sigma(t) &= \kappa(\theta - \sigma(t)) + v\sqrt{\sigma (t)} dW_2(t) + dJ^{\sigma},
\end{align*}
for $ 0\leqslant t \leqslant T $ and with $ S(0), \sigma(0) > 0  $. Here, $ \mu_S = r - \lambda\xi_S $ is the drift rate, where $r\geqslant0$ is the risk-free interest rate. The two-dimensional jump process $ (J^S,J^{\sigma}) $ is a compound Poisson process with intensity $ \lambda\geqslant0$. The distribution of the jump size in variance is assumed to be exponential with mean
$\upsilon$. In respect to jump size $z^{\sigma}$ in the variance process, $J+1$ has a log-normal distribution $p(z^S,z^{\sigma})$ with the mean in $\log z^s$ being $\gamma +\rho_J z^{\sigma}$, i.e.\ the probability density function is given by
 \[ p(z^S,z^{\sigma})=\frac{1}{\sqrt{2\pi}z^S\delta \upsilon} e^{-\frac{z^{\sigma}}{\upsilon}-\frac{(\log z^S-\gamma-\rho_J z^{\sigma})^2}{2\delta^2}}.\]
The parameter $\xi_s$ is defined by $\xi_s=e^{\gamma+\frac{\delta^2}{2}}(1-\upsilon \rho_J)^{-1}-1$, where $\rho_J$ defines the correlation between jumps in returns and variance, $\gamma$ is the jump size log-mean and $\delta^2$ is the jump size log-variance. The variance has mean level $\theta$, $\kappa$ is the rate of reversion back to mean level of $\sigma$ and $v$ is the volatility of the variance $\sigma$. The two Wiener processes $W_1$ and $W_2$ have constant correlation $\rho$.

\subsection{Partial Integro-Differential Equation}
By standard derivative pricing arguments for the SVCJ model,
obtain the PIDE
\begin{eqnarray*}
\frac{\partial V}{\partial t} +
  \frac{1}{2}S^2\sigma\frac{\partial^2 V}{\partial S^2}+\rho v\sigma
  S\frac{\partial^2 V}{\partial S \partial \sigma}
  +\frac{1}{2}v^2\sigma \frac{\partial^2 V}{\partial \sigma^2} +
  (r-\lambda\xi_s)S\frac{\partial V}{\partial S} + \kappa(\theta -
  \sigma) \frac{\partial V}{\partial \sigma}\\ - (r+\lambda)V +
  \lambda \int_0^{+\infty} \int_0^{+\infty} \! V(S . z^S,\sigma+z^{\sigma},t)p(z^S,z^{\sigma}) \,
  \mathrm{d}z^{\sigma}\mathrm{d}z^{S},
\end{eqnarray*}
which has to be solved for $S,\sigma > 0$, $0 \leq t < T $ and subject
to a suitable final condition, e.g.\ $V(S,\sigma,T) = \max(K-S,0), $
in the case of a European put option, with $K$ denoting the strike price.

\noindent

Through the following transformation of variables
\begin{equation*} x=\log S , \quad \tau = T-t , \quad   y=\sigma/v \quad \text{and} \quad u= \exp (r+\lambda)V  \end{equation*}
we obtain

\begin{multline} 
\label{eq:Ptransf}
u_\tau = \frac{1}{2}vy\left(\frac{\partial^2 u
    }{\partial x^2}+\frac{\partial^2 u}{\partial y^2}\right)+\rho
  vy\frac{\partial^2 u}{\partial x \partial y}
  -\left(\frac{1}{2}vy-r+\lambda\xi_s\right)\frac{\partial u}{\partial
    x} \\+ \kappa \frac{(\theta - vy) }{v}\frac{\partial u}{\partial y} +
 \lambda \int^{+\infty}_{-\infty} \int_0^{+\infty} \tilde u(x + z^x , y + z^y , \tau ) \tilde p(z^x,z^y) \, \mathrm{d}z^y \mathrm{d}z^x =  L_D + L_I, 
\end{multline}
which is now posed on $ \mathbb{R} \times \mathbb{R} ^+ \times (0,T), $ with

$\tilde u (x,y,\tau) = u(e^x,v y,\tau) $ and $ \tilde  p (z^x,z^y) = v e^{z^x} p(e^{z^x},z^y) $.

The problem is completed by suitable initial and boundary conditions. In the case of a European put option we have initial condition $  u(x,y ,0) = \max (1- \exp(x),0), \; x \in \mathbb{R}, \; y > 0$.

\subsection{Implicit-explicit high-order compact scheme}
\label{subsec:2}
%

For the discretisation, we replace
$\mathbb{R}$ by $[-R_1,R_1]$ and $\mathbb{R}^+$ by $[L_2,R_2]$ with
$R_1,R_2 >L_2>0$. We consider a uniform grid $ Z = \{ x_i \in
[-R_1,R_1] : x_i = i h_1 ,\; i =-N, ...  ,N\} \times \{y_j \in
[L_2,R_2] : y_j = L_2 + j h_2 ,\; j=0, ... , M\} $ consisting of
$(2N+1) \times (M+1) $ grid points with $R_1 = N h_1$ , $R_2 = L_2 + M
h_2 $ and with space step $h:=h_1=h_2$ and time step $ k $. Let $
u_{i,j}^n $ denote the approximate solution of \eqref{eq:Ptransf} in $(x_i, y_j)
$ at the time $t_n = n k$ and let $u^n = (u_{i,j}^n) $.  

For the numerical solution of the PIDE we use the implicit-explicit high-order compact (HOC) scheme
presented in \cite{DuPi17}. The implicit-explicit discretisation in time
is accomplished through an adaptation of the Crank-Nicholson method
for which we shall define an explicit treatment for the two-dimensional integral operator, $L_I$. 

We refer to \cite{DuPi17} for
the details of the derivation of the finite difference scheme for the differential operator $L_D$ and the implementation of initial and boundary conditions. To 
form the SVCJ model the coefficients are adjusted, with constant $\xi_s$ replacing $\xi_B$. 

\subsection{Integral operator}

\noindent
After the initial transformation of variables we have the integral operator in the following form,
\[ L_I = \lambda \int^{+\infty}_{-\infty} \int^{+\infty}_{0}  \tilde u(x + z^x , y + z^y , \tau ) \tilde p(z^x, z^y) \,\mathrm{d}z^y \mathrm{d}z^x , \]
We make a final change of variables $ \zeta = x+z^x $ and $ \eta = y + z^y $, with the intention of studying the value of the integral at the point  $ (x_i,y_j) $, 
\begin{equation} 
\label{eq:int1}
I =   \int^{+\infty}_{-\infty} \int^{+\infty}_{0} \tilde u( \zeta , \eta , \tau ) \tilde
p(\zeta - x_i, \eta - y_j) \, \mathrm{d}\eta \mathrm{d}\zeta 
\end{equation}

\noindent
We numerically approximate the value of \eqref{eq:int1} over the rectangle $(-R_1,R_1) \times (L_2,R_2) $, with these values chosen experimentally. 

\begin{multline} 
\label{eq:int2}
I_{i,j} =   \int^{+\infty}_{-\infty} \int^{+\infty}_{0} \tilde u( \zeta , \eta , \tau ) \tilde
p(\zeta - x_i, \eta - y_j) \, \mathrm{d}\eta \mathrm{d}\zeta \\ \approx
\int^{R_1}_{-R_1} \int^{R_2}_{L_2} \tilde u( \zeta , \eta , \tau ) \tilde
p(\zeta - x_i, \eta - y_j) \, \mathrm{d}\eta \mathrm{d}\zeta
\end{multline}


\noindent To estimate the integral we require a numerical integration method of
high order to match our finite difference scheme. We choose to use the two dimensional composite Simpson's rule. With $f$ representing the integral in \eqref{eq:int2}, we have error bounded by
\[ \frac{h^4}{180}(R_2-L_2)(2R_1) \max_{\zeta \in [-R_1,R_1], \eta \in [L_2,R_2]} |f^{(4)}(\zeta,\eta)| .\]

We evaluate the integral in \eqref{eq:int2} using the two-dimensional Simpsons rule on a equidistant grid in $x,y$ with spacing $ \Delta x = \Delta y $ and $m_x$ grid-points in $(-R_1 , R_1 ) , (L_2 , R_2 ) $, where each interval has length mesh-size $h/2$. We choose $R_1, L_2$ and $R_2$ such that the value of terms on the boundary can be considered negligible. Hence,

\begin{multline*} 
I_{i,j}  \approx \frac{h^2}{36} \Bigg[ 16 \sum_{l=1}^\frac{m_x}{2} \left( \sum_{k=1}^\frac{m_x}{2} \tilde u( x_{2k-1} , y_{2l-1} , \tau ) \tilde p(x_{2k-1}-x_i, y_{2l-1}- y_j) \right) \\ + 4 \sum_{l=1}^\frac{m_x-1}{2}  \left( \sum_{k=1}^\frac{m_x-1}{2} \tilde u( x_{2k} , y_{2l} , \tau ) \tilde p(x_{2k}-x_i, y_{2l}- y_j) \right) \\ + 8 \sum_{l=1}^\frac{m_x-1}{2} \left( \sum_{k=1}^\frac{m_x}{2} \tilde u( x_{2k-1} , y_{2l} , \tau ) \tilde p(x_{2k-1}-x_i, y_{2l}- y_j) \right) \\ +8 \sum_{l=1}^\frac{m_x}{2} \left( \sum_{k=1}^\frac{m_x-1}{2} \tilde u( x_{2k} , y_{2l-1} , \tau ) \tilde p(x_{2k}-x_i, y_{2l-1}- y_j) \right) \Bigg].
\end{multline*}

\noindent
To avoid the construction of a dense matrix we compute this integral, as a product of the sums, at each time step.

If not mentioned otherwise, we use the following default parameters in
our numerical experiments: $\kappa=2$, $\theta=0.01$, $v=0.25$, $\rho=-0.5$,
$\upsilon=0.2$, $r=0.05$, $\lambda=0.2$, $\gamma=-0.5$, $\rho_J=-0.5$, $\delta^2=0.16$.

\section{Numerical Experiments}
\label{sec:3}

We perform numerical studies to evaluate the rate of convergence and computational efficiency of the scheme. For comparison we include the results for a second-order central finite difference scheme, with the use of an appropriate two-dimensional trapezoidal rule to complete the numerical integration and the inclusion of a Rannacher-style start up to combat stability issues.

\subsection{Numerical convergence}
\label{sec:numconv}

\noindent For our convergence study we refer to both the $ l_2 $-error $\epsilon_2$ and the $ l_{\infty} $-error $\epsilon_\infty $ with respect to a numerical reference solution on a fine grid with $h_{\text{ref}} =0.025. $ With the parabolic mesh ratio $k/h^2$ fixed to a constant value we expect these errors to converge as $ \epsilon= Ch^m $ for some $m$ and $C$ which represent constants. From this we generate a double-logarithmic plot $\epsilon$ against $h$ which should be asymptotic to a straight line with slope $m$, thereby giving a method for experimentally determining the order of the scheme. 

\begin{figure}[H]
	\begin{minipage}{0.45\textwidth}
	\centering
	\epsfig{file=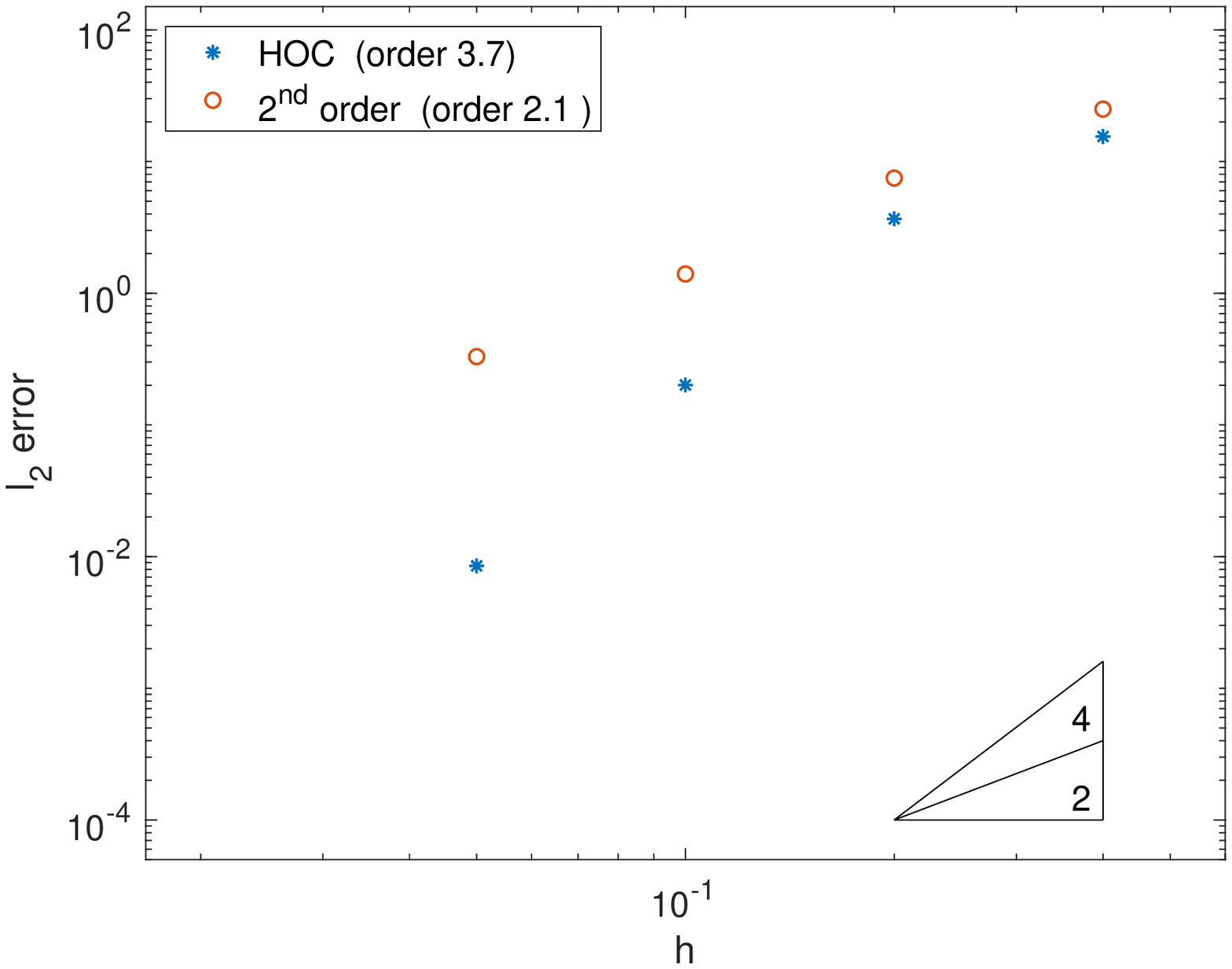, width=4.6cm}
	\caption{ $l_2$ and $l_\infty$ error in option price taken at mesh-sizes
          $h=0.4,0.2,0.1,0.05.$}
         \label{f:l2linf}
	\end{minipage}
	\begin{minipage}{0.45\textwidth}
	\centering
	\epsfig{file=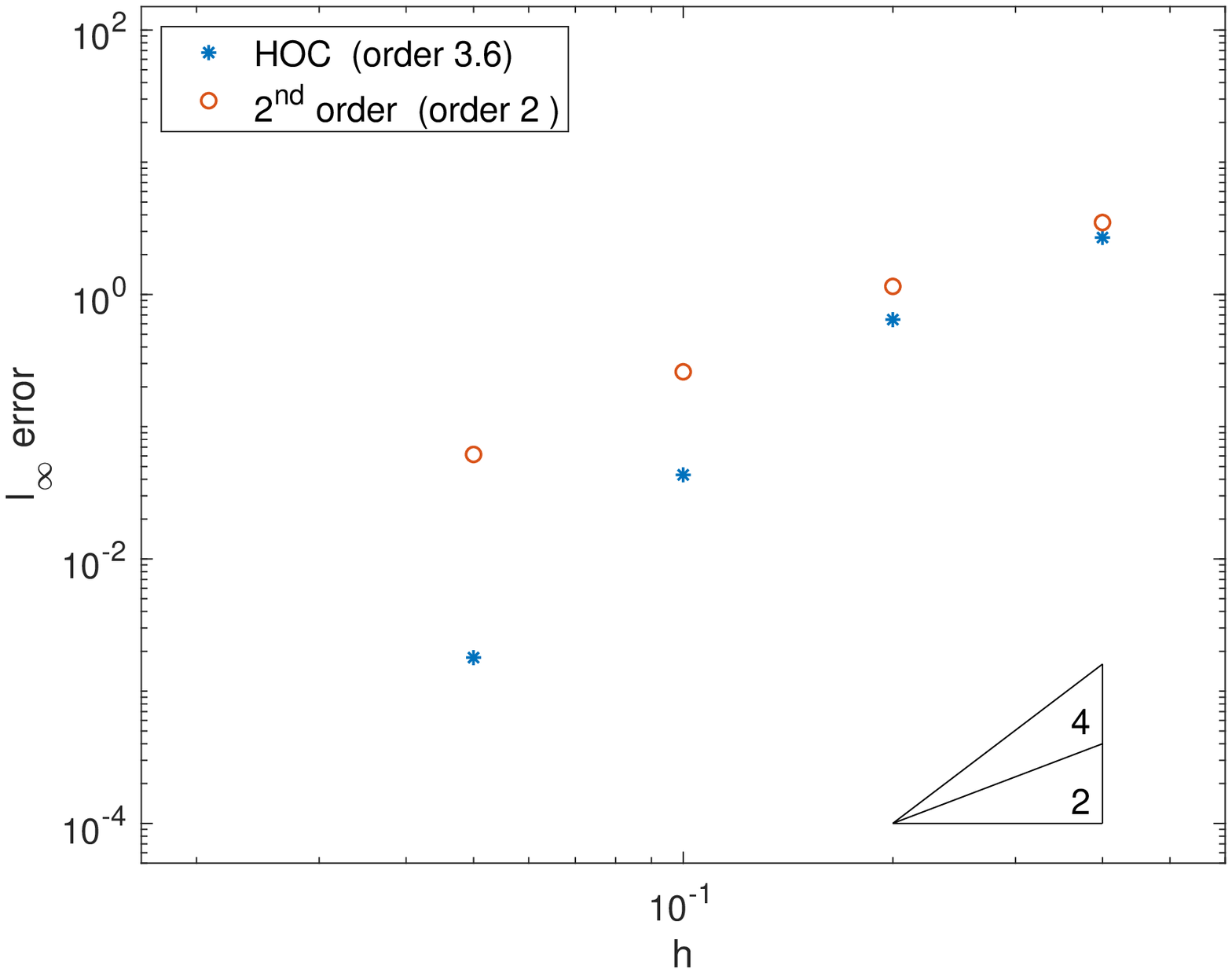, width=4.6cm}
	\end{minipage}
\end{figure}

The numerical convergence results are
included in Figure~\ref{f:l2linf}. We observe that the numerical convergence orders reflect the
theoretical order of the schemes, with the new high-order compact
scheme achieving convergence rates near fourth order.

\vspace{-0.5cm}

\subsection{Computational efficiency comparison}

\noindent 
We compare the computational time of the two schemes, looking at the time to obtain a given accuracy, taking
into account matrix setups, factorisation and boundary condition
evaluation. The timings depend obviously on technical details of the
computer as well as on specifics of the implementation, for which care was taken to
avoid unnecessary bias in the results.
All results were computed on the same laptop computer (2015 MacBook Air 11''). 

The results are shown below in Figure~\ref{f:eff}. The mesh-sizes
used for this comparison are $h=0.4, 0.2, 0.1$ and $0.05$,
with the reference mesh-size used being $h_{\mathrm{ref}}=0.025$. 

The HOC scheme achieves higher accuracy at all mesh sizes, however, this is at the expense of computation time. We attribute 
this increase to the extra computational cost associated with the Simpson's rule as compared to the trapezoidal rule.

We include the results previously seen for the Bates model, \cite{DuPi17}, to indicate the increase in 
computation time between the two models. With access to higher memory allocation it may be possible
to reduce this increase, through use of a circulant matrix and Fourier transforms to complete the numerical
integration, \cite{Salmi14}. However, it is not clear how this would be implemented with the different weightings 
assigned by Simpson's rule. 

\begin{figure}[H]
\sidecaption
\epsfig{file=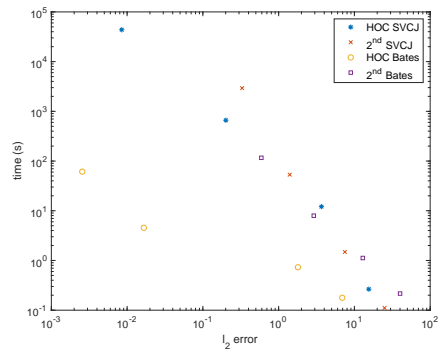, width=6.5cm}
%
%
\caption{Computational efficiency comparison taken at mesh-sizes
          $h=0.4,0.2,0.1,0.05$.}
\label{f:eff}       
\end{figure}

\begin{acknowledgement}
BD acknowledges partial support by the Leverhulme Trust research project grant `Novel discretisations for higher-order nonlinear PDE' (RPG-2015-69). 
AP has been supported by a studentship under the EPSRC Doctoral
Training Partnership (DTP) scheme (grant number EP/M506667/1).
\end{acknowledgement}

%

%
%
%

\end{document}